\documentclass[useAMS,usenatbib]{mn2e}

\usepackage{times}
\usepackage{amssymb, amsmath}
\usepackage{graphicx,rotating,amssymb,supertabular,longtable,lscape}
\usepackage{epsfig}
\usepackage[dvipdfm,colorlinks=true,linkcolor=blue,citecolor=blue,
urlcolor=blue]{hyperref}
\usepackage{enumerate}

\def\apj{\mbox{ApJ}}
\def\apjl{\mbox{ApJL}}
\def\apjs{\mbox{ApJS}}

\def\mnras{\mbox{MNRAS}}
\def\aj{\mbox{AJ}}

\def\nat{\mbox{Nature}}
\def\aap{\mbox{A\&A}}
\def\na{\mbox{NewA}}

\title[Gravitational microlensing of AGN dusty tori]{Gravitational
microlensing of AGN dusty tori}

\author[Stalevski et al.]{Marko Stalevski$^{1,2,3}$\thanks{E-mail:
mstalevski@aob.rs (MS)},
Predrag Jovanovi\'{c}$^{1,3}$, Luka \v{C}. Popovi\'{c}$^{1,3}$
\newauthor and Maarten Baes$^{2}$
\\
$^{1}$Astronomical Observatory, Volgina 7, 11060 Belgrade, Serbia\\
$^{2}$Sterrenkundig Observatorium, Universiteit Gent, Krijgslaan 281-S9, Gent, 9000, Belgium\\
$^{3}$Isaac Newton Institute of Chile, Yugoslavia Branch, Volgina 7,
11060 Belgrade, Serbia}

\begin{document}

\date{\today}

\pagerange{\pageref{firstpage}--\pageref{lastpage}} \pubyear{2012}

\maketitle

\label{firstpage}

\begin{abstract}
  We investigated gravitational microlensing of active galactic nuclei
dusty tori in the
  case of lensed quasars in the infrared domain. The dusty torus is
  modeled as a clumpy two-phase medium. To obtain spectral energy
  distributions and images of tori at different wavelengths, we used
  the 3D Monte Carlo radiative transfer code \textsc{skirt}.  A
  ray-shooting technique has been used to calculate microlensing
  magnification maps. We simulated microlensing by the
  stars in the lens galaxy for different configurations of the lensed
  system and different values of the torus parameters, in order to
  estimate (a)\ amplitudes and timescales of high magnification
  events, and (b)\ the influence of geometrical and physical
  properties of dusty tori on light curves in the infrared domain. We
  found that, despite their large size, dusty tori could be
  significantly affected by microlensing in some cases, especially in
  the near-infrared domain (rest-frame). The very long time-scales of
  such events, in the range from several decades to hundreds of years,
  are limiting the practical use of this method to study the
  properties of dusty tori . However, our results indicate that,
  when studying flux ratios between the images in different wavebands
  of lensed quasars, one should not disregard the possibility that the
  near and mid-infrared flux ratios could be under the influence of
  microlensing.
\end{abstract}

\section{Introduction}
Gravitationally lensed systems with multiple images represent a
powerful tool to study the structure of both the galaxy which acts as
the lens and the background source. In a number of lensed systems in
which a quasar is the source, the flux ratios between the lensed
images deviate from those predicted by the simple lens models
\citep[see e.g.][]{koch91,keet03,gold10}. The fluxes, in different
wavebands, can be contaminated by different effects such as
microlensing by the stars \citep[e.g.][]{schwam90} or millilensing by
a massive structure in the lens galaxy \citep[e.g.][]{maosch98}, dust
extinction \citep[e.g.][]{elis06} and also by the time delay itself
\citep[e.g.][]{pc05}. Consequently, the flux ratio anomaly
observed in some lensed quasars can be caused by extinction and/or
gravitational microlensing/millilensing \citep[see
e.g.][]{pc05,yonehara08}.

The size of the source has a large effect on the fluctuations
due to
microlensing. As a large extended source covers a
larger area of a microlensing magnification pattern in the source
plane at any given time than a small source, its brightness varies
less as it moves relative to the lens and observer
\citep{mortonson05}. As a general rule, the variability of a lensed
source is significantly affected by microlensing only if the source
is smaller than the relevant length scale - the projection of the
Einstein radius of a microlens on to the
source plane \citep{courbin02}. Since the sizes of different
emitting regions of quasar are wavelength-dependent, microlensing by
the stars in the lens galaxy will lead to a wavelength-dependent
magnification. The X-ray radiation is coming from the very compact
region in the innermost part of the accretion disk, and therefore,
it will be magnified more than the radiation in the UV and optical
bands, coming from outer, larger parts of the disk.
Thus, although the phenomenon of gravitational lensing is achromatic,
due to the complex structure of emission regions, ``chromatic''
effect may arise in a lensed quasar system 
\citep[see e.g.][]{pc05,jovanovic08,mo09,mo11}. The ``chromaticity''
in lensing effect can be used to investigate both, an unresolved
structure of the innermost region of quasars \citep[see
e.g.][]{wyithe02,aba02,pop03,pop06,bate08,mo09,dai10,black11,
garsden11} and the structure of the lens galaxy \citep[see
e.g.][]{ic05,chiba05,xu10}. Moreover, comparing flux ratios at
different wavelengths makes it possible to constrain the amount of
micro- and milli-lensing present in the system, and the sizes of the
perturbers \citep[see e.g.][]{gold10}. 

Since the X-ray and UV/optical radiations
are coming from relatively compact regions (from several light-days to
a light-month), the flux ratios in these wavebands are
sensitive to both microlensing by the stars and millilensing by a cold
dark matter (CDM) substructure \citep[see
e.g.][]{mm01,pc05,dk06,jovanovic08,gold10,xu10}. On the other hand,
the infrared (IR) emitting region of a quasar is assumed to be a
toroidal structure of dust, with dimensions significantly larger than
the the projection of the Einstein radius of a microlens on to the
source plane. Therefore one would
expect that the IR radiation of lensed quasars would only be affected
by the relatively massive subhalos (millilensing)
\citep[see][]{ic05,chiba05,sluse06,yonehara08,min09,xu10,fk11}.
However, certain geometrical and physical properties of the dusty
torus can conspire to allow non-negligible microlensing effects in the
infrared domain.

Infrared spectra of most quasars are dominated by thermal emission
from hot dust in their tori, or alternatively, by nonthermal
synchrotron emission from the regions near their central black holes
\citep{agol00}. Variability in the infrared band due to gravitational
microlensing could be used to constrain the size of the infrared
emission region, and hence to distinguish between the thermal and
synchrotron mechanisms. If the infrared radiation varies on timescales
shorter than decades, then its emission region is smaller, located
closer to the central black hole, and its emission is nonthermal,
while longer timescales indicate a larger, thermal region
\citep{neugebauer99}. Additionally, chromatic effects in the infrared
band have been observed in some of the lensed quasars, where
the color differences between their multiple images were detected
\citep{yonehara08}. The most realistic scenario that can explain the
observed color differences is gravitational microlensing, in contrast
to the dust extinction and the intrinsic variability of quasars
\citep{yonehara08}.

Some previous theoretical and observational studies suggested that the
infrared emission of quasars is not significantly affected by
microlensing, implying that it is most likely produced in their dusty
tori. For instance, \citet{agol00} studied the mid-IR emission of
Q2237+0305 observed by Keck and found that it was not affected by
microlensing, which ruled out the synchrotron mechanism and supported
the model with hot dust extended on a length scale of more than 0.03
pc. \citet{wyithe02} used mid-IR and $V$-band flux ratios for images
$A$ and $B$ of Q2237+0305 to infer the size of the mid-IR emission
region and found that it was comparable to or larger than the Einstein
Ring Radius (ERR) of the microlens, and hence at least two orders of
magnitude larger than the optical emission region. They used simple
Gaussian and annular intensity profiles of the dusty torus and found
that the results were dependent on the assumed source profile
\citep{wyithe02}. Recent Spitzer observations of the same
gravitationally lensed quasar \citep{agol09} showed that a dusty torus
model with a small opening angle could satisfactorily explain the
shape of the observed infrared SED, excluding an offset in wavelength
of the silicate feature. However, the same authors found that the
near-IR fluxes are increasingly affected by microlensing toward
shorter wavelengths and that this wavelength dependence is consistent
with a model in which a dusty torus and an accretion disk both
contribute to the infrared radiation near 1 $\mathrm{\mu m}$
\citep{agol09}.

In this paper we present simulations of gravitational microlensing of
active galactic nucleus (AGN) dusty tori in the infrared domain. We
consider microlensing by
stars in the lens galaxies, in the case of lensed quasar systems. We
modeled the dusty torus as a clumpy two-phase medium. To obtain
spectral energy distributions and images of the torus at different
wavelengths, we used the 3D radiative transfer code
\textsc{skirt}. For generating microlensing magnification maps, a
ray-shooting method was used. To take into account the size of the
dusty tori (as they are larger than the Einstein ring radius of the
microlens projected on the source plane), the simulated images of the
tori are convolved with the magnification maps. We simulated
microlensing magnification events for different configurations of the
lensed system and different values of the torus parameters. The aims
of this paper are to estimate (a)\ amplitudes and timescales of high
magnification events, and (b)\ the influence of geometrical and
physical properties of dusty tori on microlensing light curves in the
infrared domain.

The paper is organized as follows. In Section 2 we give a description
of our dusty torus model, the method we used to calculate
microlensing magnification map, and the parameters we adopted in this
study. In Section 3 we present and discuss the results of simulated
microlensing light curves of the dusty torus. In Section 4 we outline
our conclusions.
\section{Model}
\label{sec:mod}
\subsection{The radiative transfer code \textsc{skirt}}

We have used the radiative transfer code \textsc{skirt} to calculate
spectral energy distributions (SED) and images of torus at different
wavelengths. \textsc{skirt} is a 3D Monte Carlo radiative transfer
code, initially developed to investigate the effects of
dust extinction on the photometry and kinematics of galaxies
\citep{baes03}. Over the years, the code evolved into a flexible tool
that can model the dust extinction, including both absorption and
scattering, and the thermal re-emission of dust, under the assumption
of local thermal equilibrium (LTE) \citep{baes05a, baes05b}. This LTE
version of \textsc{skirt} has been used to model different
environments, such as, circumstellar disks \citep{vidal07}, clumpy
tori around active galactic nuclei \citep{stalevski12} and a variety
of galaxy types \citep{baes10,dlooze10}. Recently, the code was
adapted to include the emission from very small grains and polycyclic
aromatic hydrocarbon molecules \citep{baes11}.

\subsection{Dusty torus model}
\label{sec:tor}
According to the AGN unification model, the central continuum source
(accretion disk) is surrounded by the geometrically and optically
thick toroidal
structure of dust and gas with an equatorial visual optical depth
much larger than unity. In order to prevent the dust grains from
being destroyed by the hot surrounding gas, \citet{krolikbegel88}
suggested that the dust in torus is organized in a large
number of optically thick clumps. This dusty torus absorbs the
incoming radiation and re-emits it, mostly in the infrared domain.

The details of the model of torus we used in this work can be found
in \citet{stalevski12}; here we will present only some of its
properties relevant for this study. We modeled the torus as a 3D
two-phase medium with high-density clumps and low density medium
filling the space between the clumps. 

We approximate the obscuring toroidal dusty structure with a conical
torus (i.e. a flared disk). Its characteristics are defined by (a)\
half opening angle $\Theta$, (b)\ the inner and outer radius,
$R_{\text{in}}$ and $R_{\text{out}}$ respectively, and (c)\ the
parameters $p$ and $q$ that describe the dust density
distribution. The inner radius is calculated according to the
prescription given by \citet{barvainis87},
\begin{equation}\label{eqn:rin}
  \left(\frac{R_{\text{in}}}{\text{pc}}\right)
  \simeq
  1.3
  \left(\frac{L_{\text{AGN}}}{10^{46}~{\text{erg}}\,{\text{s}}^{-1}}\right)^{0.5}
  \left(\frac{T_{\text{sub}}}{1500~{\text{K}}}\right)^{-2.8}
\end{equation}
where $L_{\text{AGN}}$ is the bolometric ultraviolet/optical
luminosity
emitted by the central source and $T_{\text{sub}}$ is the sublimation
temperature of the dust grains.

We describe the spatial distribution of the dust density with a law
that allows a density gradient along the radial direction and with
polar angle, as the one adopted by \citet{granatodanese94}:
\begin{equation}\label{eqn:dens}
\rho\left(r,\theta \right)\propto r^{-p}e^{-q|cos\theta|} ,
\end{equation}
where $r$ and $\theta$ are coordinates in the adopted coordinate
system. The dust mixture consists of separate populations of graphite
and silicate dust grains with a classical MRN size distribution
\citep*{mrn77}. The total amount of dust is fixed based on the
equatorial optical depth at $9.7$ $\mu$m ($\tau_{9.7}$). Dust is
distributed on a 3D Cartesian grid composed of a large number of
cubic cells. To generate a clumpy, two-phase medium, we apply the
algorithm described by \citet{wittgordon96}. The larger clumps are
formed by forcing high-density state into several adjoining cells.
\subsubsection{Spectral energy distribution of the primary continuum
source}
The primary continuum source of dust heating is the intense
UV-optical continuum coming from the accretion disk. A good
approximation of its emission is a central, point-like energy source,
emitting isotropically. Its SED is very well approximated by a
composition of power laws with different spectral indices in
different spectral ranges. The adopted values are:
\begin{equation}\label{eqn:source}
\lambda L_\lambda
\propto\left\{
\begin{array}{lrr}
\lambda^{1.2}  &  \; 0.001 \leq \lambda \leq 0.01  \\
\lambda^{0}    &  \; 0.01  < \lambda \leq 0.1   \\
\lambda^{-0.5} &  \; 0.1   < \lambda \leq 5     \\
\lambda^{-3}   &  \; 5     < \lambda \leq 50   
\end{array}
\right.
\end{equation}
where $\lambda$ is expressed in $\mu$m. 
These values have been quite
commonly adopted in the literature, and come from both observational
and theoretical arguments \citep[see e.g.,][]{schartmann05}.
\subsubsection{Parameters of the dusty torus model}
\label{sec:torpar}
For the bolometric luminosity of the primary source (the accretion
disk) we adopted the value of $10^{12}\ L_{\odot}$
\citep[e.g.][]{davislaor11}.  According to the Eq.~(\ref{eqn:rin}),
assuming a dust sublimation temperature of $1500$ K, the corresponding
value of the inner radius of the torus is $R_{\text{in}}=0.8$
pc. Recent observations of nearby AGN suggest that the size of tori is
likely restricted to several parsecs \citep{alherrero11,almeida11}. It
is arguable whether this holds also for the tori in AGNs at much
larger redshifts (i.e.\ quasars).  Preliminary results from a recent
study of high redshift quasars by \citet{leipmeis12} suggest that the
hot dust properties do not change significantly with either redshift
or luminosity. Therefore, for the outer radius of torus we adopted
the values of $3$ and $10$ pc. The half opening angle of the torus
takes values of $30^\circ$, $50^\circ$ and $70^\circ$. For the dust
density distribution parameters we adopted the following values:
$p=0,0.5,1$ and $q=0,1,2$. The relative clump size, defined as the
ratio of the outer radius of the torus over the clump size, is
$\xi=12.5$. The equatorial optical depth (determining also the total
amount of dust) is fixed at $\tau_{9.7}=5$. We calculated each model
at three inclinations, $i=0^\circ, 70^\circ, 90^\circ$, where
$i=0^\circ$ corresponds to face-on view (type 1 AGN), and
$i=90^\circ$ to edge-on view (type 2 AGN). The emission for all
models is calculated on an equally spaced logarithmic
wavelength grid ranging from $0.001$ to $1000\ \mu$m.
\subsection{Microlensing model}
\label{sec:mlmod}
Different mirolensing models can be used for explaining the
observed microlensing phenomena in quasars, such as microlensing by an
isolated compact object \citep{chang79,chang84}, or by a number
of microdeflectors located within an extended object -- the model called
''straight-fold caustic'' \citep{schneider92}. However, the most realistic model
is the so-called microlensing map (pattern) or caustic network which
is usually
applied in order to obtain a spatial distribution of magnifications in the
source plane (where the dusty torus of quasar is located), produced by
randomly distributed stars in the lensing galaxy \citep[for more details see
e.g.][]{schneider92}. The lensing galaxy can be assumed as
thin when compared with the whole light path, since its extent in
the direction of the optical axis is much smaller than the angular diameter
distances between observer and lens, observer and source and
lens and source. In such case the dimensionless lens equation
reads \citep[][]{schneider92}:
\begin{equation}\label{alpha}
\vec{y}=\vec{x}-\vec{\alpha}\left(\vec{x}\right) ,
\end{equation}
where $\vec{x}$ and $\vec{y}$ are normalized image and source
positions, respectively, and $\vec{\alpha}\left(\vec{x}\right)$ is the
deflection angle due to light banding in the gravitational field of the lens.
The deflection potential is then given by:
\begin{equation}\label{psi0}
\psi\left(\vec{x}\right)=\frac{1}{\pi}\int\kappa\left(\vec{x'}
\right)\ln\vert\vec{x}-\vec{x'}\vert d^2 x',
\end{equation}
and it is related to the deflection angle by:
$\vec{\alpha}\left(\vec{x}\right)=\nabla\psi\left(\vec{x}\right)$ and to the
dimensionless surface mass density $\kappa$ by the 2-dimensional Poisson
equation: $\nabla^2\psi=2\kappa$. For a field of point masses with an external
shear $\gamma$ and a smooth mass distribution $\kappa_c$, the normalized lens
equation (\ref{alpha}) is usually written as \citep{kayser86}:
\begin{equation}\label{lenseq}
\vec y = \sum\limits_{i = 1}^N {m_i \frac{{\vec x - \vec x_i
}}{{\left| {\vec x - \vec x_i } \right|^2 }}} + \left[
{\begin{array}{*{20}c} {1 - \kappa_c  + \gamma } & 0  \\ 0 & {1 -
\kappa_c  - \gamma }  \\ \end{array}} \right]\vec x,
\end{equation}
where the sum describes light deflection by the stars and the last term
is a quadrupole contribution from the lensing galaxy. The total surface mass
density or convergence can be written as $\kappa=\kappa_\ast+\kappa_c$, where
$\kappa_\ast$ represents the contribution from the compact microlenses. The
magnification map for some specific microlensing event can be generated if the
following two parameters are known: the convergence - $\kappa$, and the shear
due to the external mass - $\gamma$.

For generating microlensing magnification maps we used a ray-shooting
method \citep{kayser86, schneider86, schneider87, wambsganss90}. In this method,
rays are followed backwards from the observer through
the lens plane, to the source plane. First, we generate a random
star field in the lens plane based on the parameter $\kappa$. After
that, we solve the Poisson equation $\nabla^2\psi=2\kappa$ in the
lens plane numerically, so we can determine the lens potential $\psi$
in every point of the grid in the lens plane. To solve the Poisson
equation numerically one has to write its finite difference form:
\begin{equation}\label{psi}
\psi_{i+1,j}+\psi_{i-1,j}+\psi_{i,j+1}+\psi_{i,j-1}-4\psi_{i,j}
=2\kappa_{i,j}.
\end{equation}
Here we used the standard 5-point formula for the two-dimensional
Laplacian. The next step is the inversion of equation (\ref{psi})
using
Fourier transforms. After some transformations we obtain:
\begin{equation}\label{hatpsi}
\hat{\psi}=\frac{\hat{\kappa}_{mn}}{2(\cos{\frac{m\pi}{N_{1}}}+\cos{
\frac{n\pi}{N_{2}}}-2)},
\end{equation}
where $N_1$ and $N_2$ are dimensions of the grid in the lens plane.
Now, using the finite difference technique, we can compute the
deflection angle $ \vec{\alpha}=\nabla\psi $ in each point of the
grid
in the lens plane. After computing the deflection angle, we can map
the regular grid of points in the lens plane, via lens equation, onto
the source plane. These light rays are then collected in pixels in
the
source plane, and the number of rays in one pixel is proportional to
the magnification due to microlensing at this point in the source
plane. Due to the relative motion between the observer, lens and
source, the magnification over time will change, and a light curve
for a small, pointlike source can be found by tracing a path across
the magnification map.

Apart from the convergence $\kappa$ and shear $\gamma$, another input
parameter is width of the microlensing magnification map expressed in
units of the Einstein ring radius in the source plane. The Einstein
ring radius in the lens plane is defined as
\begin{equation}\label{eqn:err}
{\text{ERR}}
=
\sqrt{\dfrac{4Gm}{c^2}\,\dfrac{D_{\text{l}}D_{\text{ls}}}{D_{\text{s}
}}} ,
\end{equation}
and its projection in the source plane is:
\begin{equation}\label{eqn:re}
R_{\text{E}}
=
\dfrac{D_{\text{s}}}{D_{\text{l}}}\,{\text{ERR}}
=
\sqrt{\dfrac{4Gm}{c^2}\,\dfrac{D_{\text{s}}D_{\text{ls}}}{D_{\text{l}
}}
} ,
\end{equation}
where $G$ is the gravitational constant, $c$ is the speed of light,
$m$ is the microlens mass and $D_{\text{l}}$, $D_{\text{s}}$ and
$D_{\text{ls}}$ are the cosmological angular distances between
observer-lens, observer-source and lens-source, respectively.
\subsubsection{Parameters of the microlensing magnification map}
\label{sec:mlpar}
The microlensing magnification map is calculated for typical values
of
average surface mass density and shear, $\kappa=\gamma=0.4$. The map
has $1156\ R_{\text{E}}$ on a side. For a source fixed at redshift of
$z_{\text{s}}=2$, and for the lens at $z_{\text{l}}=0.05$ and $0.5$,
this is equal to $61.42$ and $19.99$ pc in the source plane,
respectively). The size of the map is chosen because of the large
dimensons of the dusty torus compared to $R_{\text{E}}$. But this
particular value is also chosen for numerical reasons -- the
pixel size has to be the same in the images of the torus and the
magnification map. Thus, the size and resolution of the images of the
torus and the magnification map cannot be choosen independently. A
flat cosmological model is assumed, with $\Omega_{M}=0.27$,
$\Omega_{\Lambda}=0.73$ and $H_{0}= 71\ \rm km\ s^{-1} Mpc^{-1}$. The
mass of microlens is taken to be $1 M_{\odot}$ in all simulations.
\section{Results and discussion}
\label{sec:res}
In this section, we first present the microlensing magnification map
used in this study and demonstrate the wavelength dependency of the
AGN dusty torus size. Then, the size the of torus is taken into
account by convolving the magnification map with images of torus at
different wavelengths. Finally, we discuss light curves of simulated
magnification events, the amplitudes and timescales of these events,
and the influence of different torus parameters. Note that, throughout
this paper, we always refer to the rest-frame wavelengths.

We calculated microlensing light curves for a source fixed at a
redshift of $z_{\text{s}}=2$, while the lens galaxy redshift takes
values of $z_{\text{l}}=0.05$ and $0.5$. The former value of
lens galaxy redshift ($z_{\text{l}}=0.05$) roughly replicates the
well-known lensed system Q2237+0305 (also known as the ``Einstein
cross''), which is particularly susceptible to microlensing. The
redshift of the lens in this system ($z_{\text{l}}=0.04$) is so low
that the apparent angular velocity of the microlenses, in projection
on the plane of the sky, is much higher than in other systems.
Moreover, the Einstein rings of these microlenses have a larger
angular diameter, making it more likely that they are larger than the
source \citep{courbin02}. The second value chosen for calculation of
magnification map ($z_{\text{l}}=0.5$) represents a more
typical value of lens galaxy redshift in lensed systems.

\begin{figure*}
\centering
\includegraphics[height=0.48\textwidth]{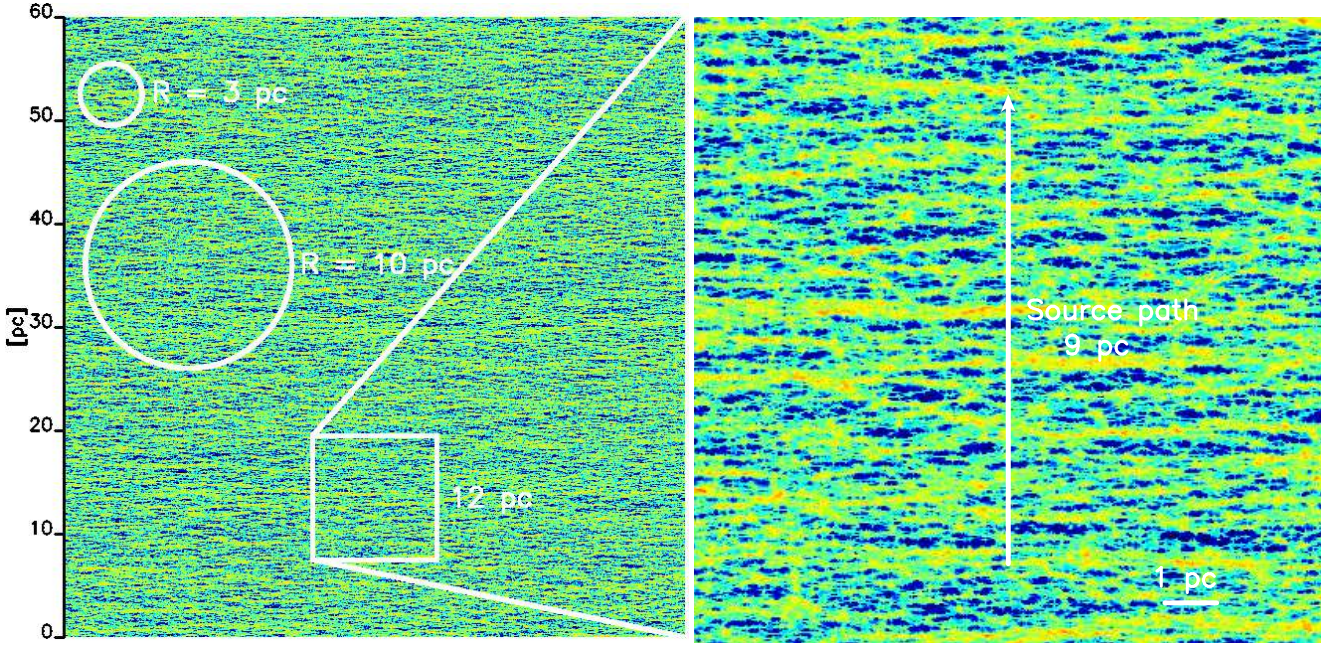}
\includegraphics[height=0.47\textwidth]{fig1b.eps}
\caption{\textit{Left panel:} Microlensing magnification map, with
$61.42$ pc ($1156\ R_{\text{E}}$) on the side. Average surface mass
density and shear take value of $\kappa=\gamma=0.4$. White circles
indicate sizes of tori used in this study ($R_{\text{out}}=3$ and
$10$ pc). \textit{Right panel:} zoom-in on the square with $12$ pc on
the side, from which the light curve of microlensing event is
extracted (vertical white line).}
\label{fig:mlmap}
\end{figure*}

The microlensing magnification map shown in Fig.~\ref{fig:mlmap} is
calculated for the case of $z_{\text{l}}=0.05$ (and other parameters
as described in Section~\ref{sec:mlpar}). The left panel represents
the whole map, with $\sim61.42$ pc ($1156\ R_{\text{E}}$) on the
side;
for comparison, the sizes of the tori used in this study
($R_{\text{out}}=3$ and $10$ pc) are indicated with two white
circles. The right panel is a zoom-in of the region with a side of
$12$ pc, from which the light curve of the microlensing events is
extracted (vertical white line). The distribution of
magnification is highly non-liner, with regions of low and high
magnification and with a maximum magnification along the sharp
caustic lines. Magnification pattern is color coded, with blue, cyan,
yellow, orange and red regions corresponding to increasingly higher
magnification. When a source crosses a caustic, a large change in
magnification is expected.
\subsection{Wavelength dependency of torus size}

\begin{figure*}
\centering
\includegraphics[height=0.245\textwidth]{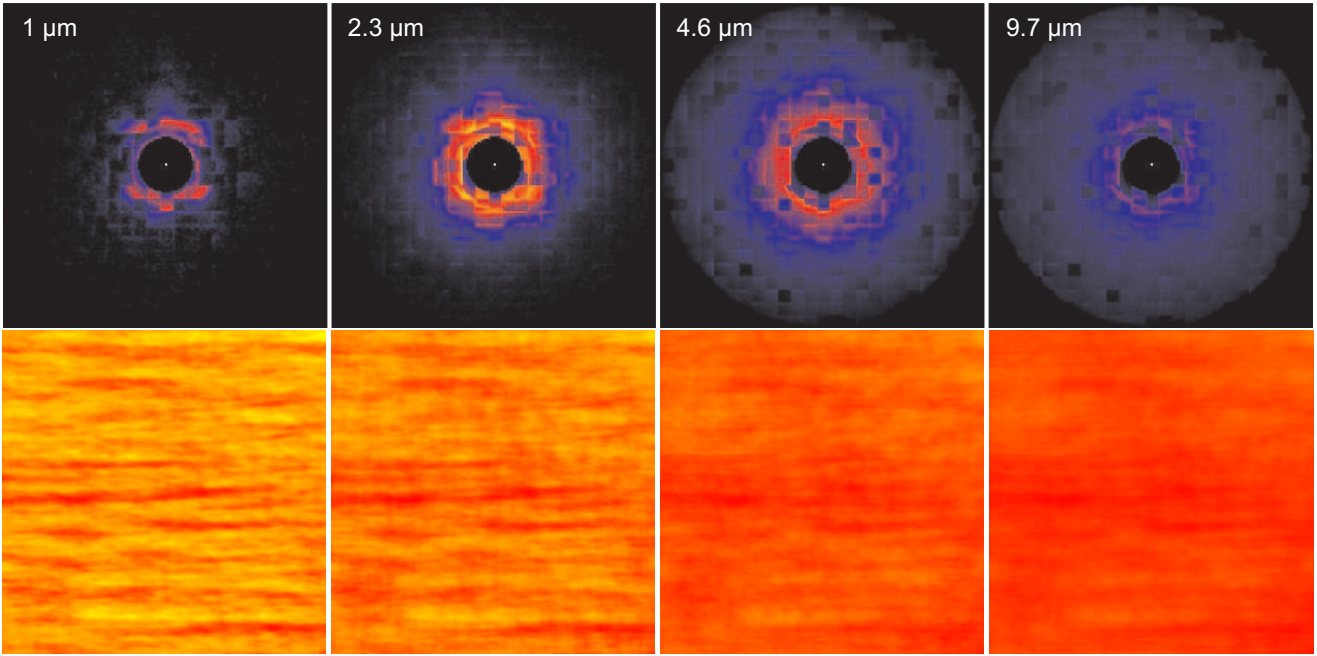}\\
\includegraphics[height=0.243\textwidth]{fig2b.eps}
\includegraphics[height=0.243\textwidth]{fig2c.eps}
\includegraphics[height=0.243\textwidth]{fig2d.eps}
\includegraphics[height=0.243\textwidth]{fig2e.eps}\\
\caption{\textit{Top row:} images of torus at different wavelengths,
face-on view. From left to right, panels represent model
images at $1$, $2.3$, $4.6$, and $9.7\ \mu$m. All images are given in
the same, logarithmic color scale. The visible squared structure is
due to the clumps which in our model are in the form of cubes. The
adopted values of torus model parameters are: optical depth
$\tau_{9.7}=5$, dust distribution parameters $p=1$ and $q=2$, the
half opening angle $\Theta=50^\circ$, the relative clump size
$\xi=12.5$, the inner and outer radius $R_{\text{in}}=0.8$ and
$R_{\text{out}}=3$ pc, respectively.
\textit{Bottom row:} microlensing magnification maps after
convolution with the corresponding tori images from the top row.
Maps correspond to the region shown in the right panel of
Fig.~\ref{fig:mlmap}. For clarity, a different scale of coloring for
each map is adopted, so that the details of each image are visible.}
\label{fig:torimg}
\end{figure*}

In the top row of Fig.~\ref{fig:torimg} we present images of the torus
at different wavelengths. At shorter wavelengths, it is the radiation
from the inner (and hotter) region that dominates. Thus, at
near-infrared wavelenghts, the torus appears much more compact
compared to its physical outer radius. On the other hand, at longer
wavelengths, the emission arises from the colder dust. As this colder
dust is placed further from the centre, all the way to the outer
radius, the torus will appear larger at longer wavelenghts.
Therefore, the size of torus is wavelength dependent, as seen in the
top row Fig.~\ref{fig:torimg}.

The microlensing magnification maps are calculated for point
sources. As the dusty torus is larger than the typical size of
microlens $R_{\text{E}}$, they have to be treated
as extended sources. To take this into account, the magnification
map shown in Fig.~\ref{fig:mlmap} is convolved with the images of the
torus at the different wavelenghts. The bottom row of
Fig.~\ref{fig:torimg} shows the magnification maps after convolution
with the corresponding torus images from the top row.
\subsection{Simulated light curves of microlensing events}
\subsubsection{Wavelength dependence}
\label{sec:lcwl}

\begin{figure*}
\centering
\includegraphics[height=0.34\textwidth]{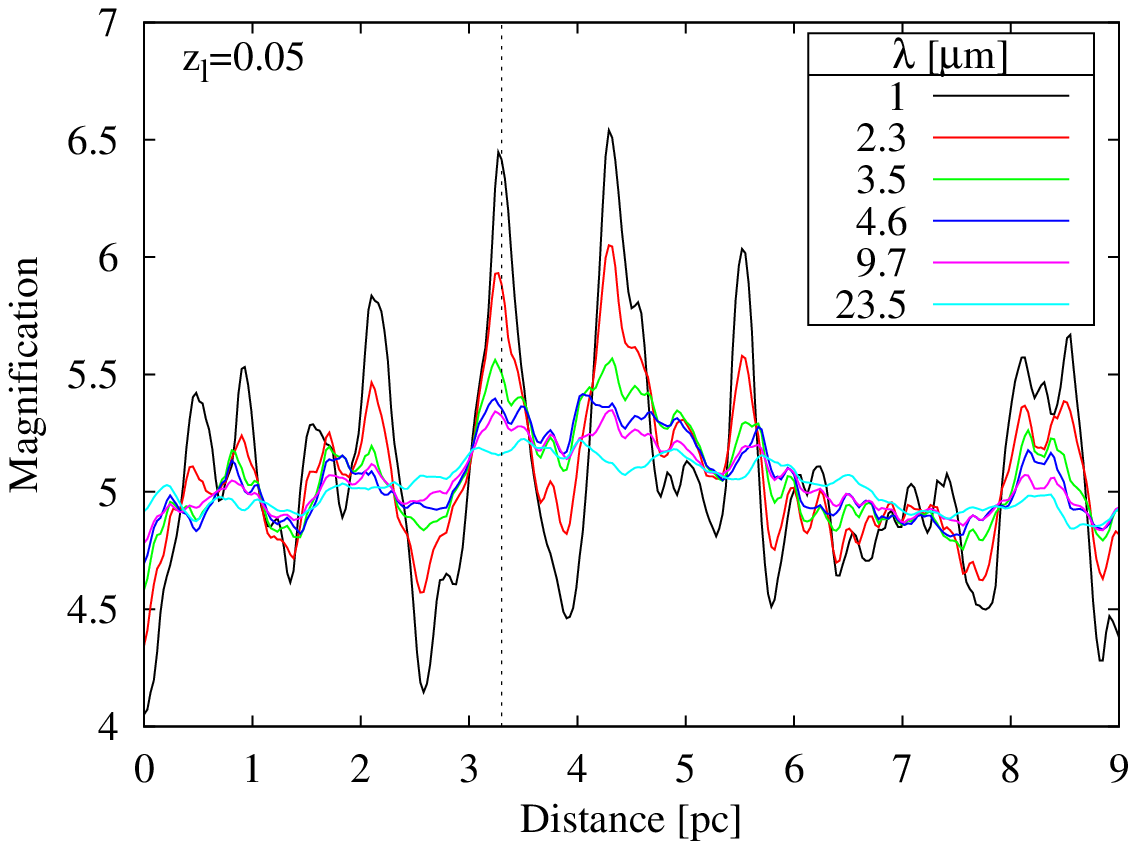}
\includegraphics[height=0.34\textwidth]{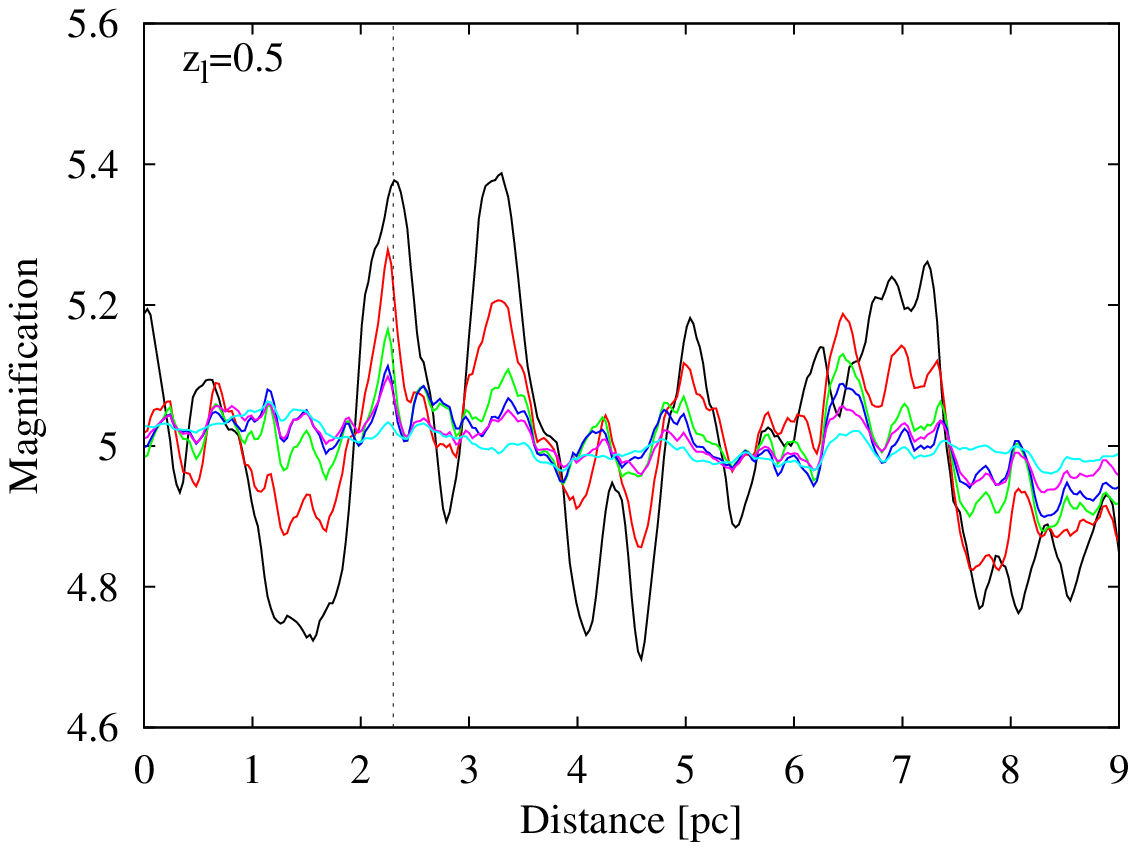}
\caption{Light curves of magnification events at different
rest-frame wavelengths (indicated in the legend),
extracted from the magnification maps convolved with the
corresponding images of tori. \textit{Left panel: $z_{\text{l}}=0.05$}.
\textit{Right panel: $z_{\text{l}}=0.5$}. Note the different range of y axis
shown in the two panels. The dotted line in both panels indicates
two HMEs referenced throughout the text. The distance given in the x
axis represents the distance crossed by the source, relative to
the caustics, for the given values of the source and lens redshifts.
The value ``0'' corresponds to the beginning of the path from which
the lightcurve was extracted.}
\label{fig:lc}
\end{figure*}

\begin{figure*}
\centering
\includegraphics[height=0.34\textwidth]{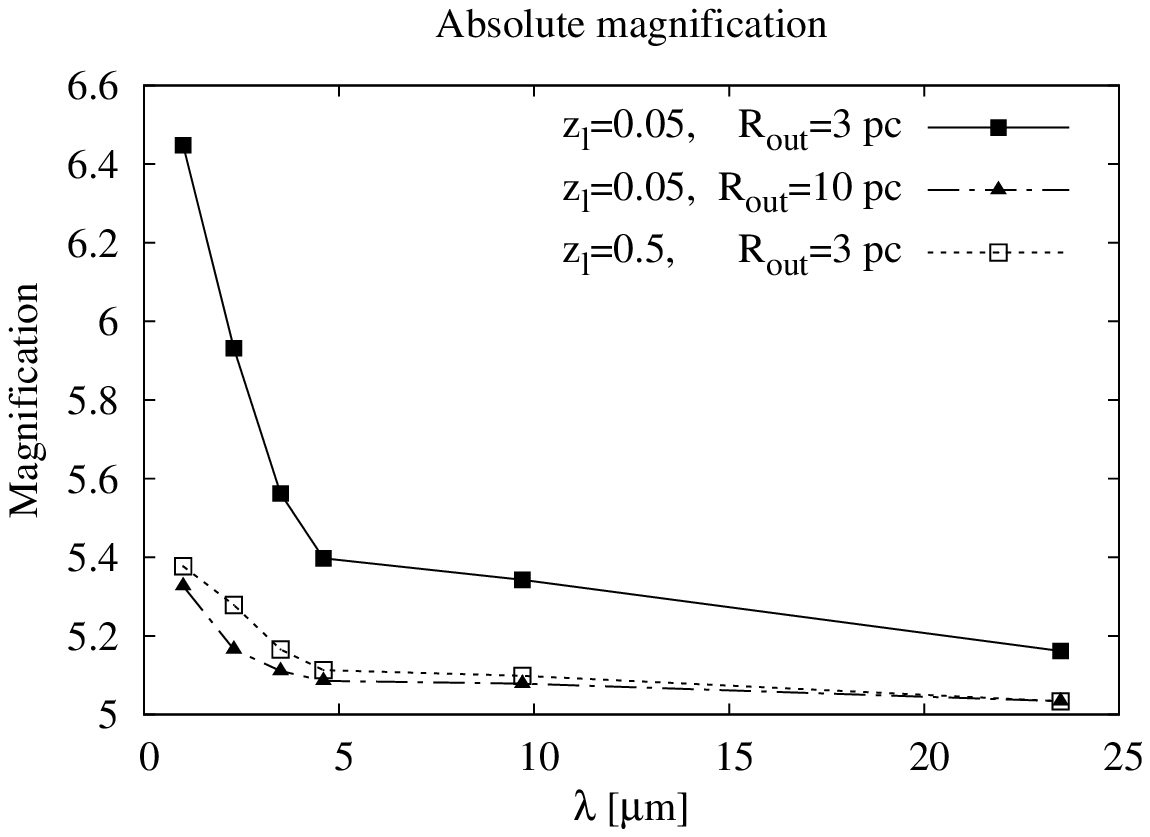}
\includegraphics[height=0.34\textwidth]{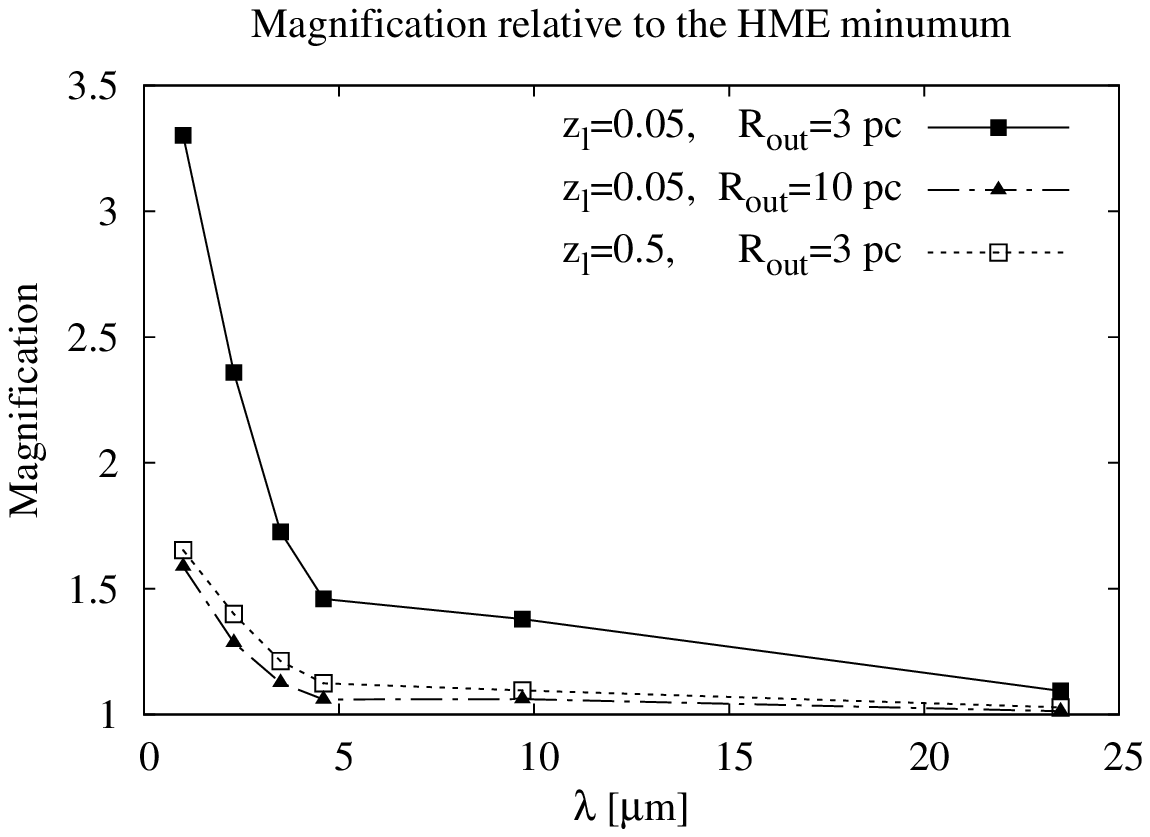}
\caption{Magnification amplitude as a function of restframe
wavelength, for
prominent HMEs seen in Fig.~\ref{fig:lc}. Full squares: HME peaking
at $\sim3.3$ pc in left panel of Fig.~\ref{fig:lc} ($z_{\text{l}}=0.05$).
Triangles: HME peaking at $\sim2.3$ pc in the right panel of
Fig.~\ref{fig:lc} ($z_{\text{l}}=0.5$). Empty squares: the same as for the
full squares, but for larger value of torus outer radius.
\textit{Left
panel:} absolute magnification amplitude. \textit{Right panel:}
magnification amplitude relative to the minimum (beginning of
the HME).}
\label{fig:mamp}
\end{figure*}

In Fig.~\ref{fig:lc} we present light curves of magnification events
at different wavelengths. The light curves are extracted from the
magnification maps convolved with the corresponding images of the
tori, along the path shown in the right panel of Fig.~\ref{fig:mlmap}
(starting from bottom, going to the top). The parameters of the torus
are the same as taken in Fig.~\ref{fig:torimg}. The left panel of
Fig.~\ref{fig:mamp} illustrates the dependence of the absolute
magnification amplitude on wavelength, for the prominent high
magnification events (HME), peaking at $\sim3.3$ and $\sim2.3$ pc
(indicated with a dotted line in the left and right panel of
Fig.~\ref{fig:lc}, respectively). From
these figures we see that, as a consequence of the wavelength
dependency of the torus size, the magnification amplitude of the
microlensing events is also wavelength dependent. The magnification is
highest at near-infrared wavelengths, decreasing toward the
mid-infrared range, and remains almost constant in the far-infrared
part of the SED. Note that, as the torus size is larger than the
typical size of microlens $R_{\text{E}}$, its radiation will always
be
magnified by a certain factor. Therefore, in the right panel of
Fig.~\ref{fig:mamp} we also illustrate the amplitudes
relative to the minimum, i.e. to the beginning of a HME.

We note here that, at near-infrared wavelengths, the accretion disk
radiation may still have a significant contribution to the SED. As
the accretion
disk is much smaller than the typical size of $R_{\text{E}}$, it can
lead to microlensing events of much higher amplitudes and shorter
timescales.  However, in this work we investigate only the
long-term variations caused by microlensing of the dusty torus;
microlensing of the accretion disk has been throughly studied in the
literature
\citep[e.g][]{jovanovic08,blackkoch10,morgan10,dexteragol11}

\subsubsection{Dependence on torus parameters}
\begin{figure*}
\centering
\includegraphics[height=0.34\textwidth]{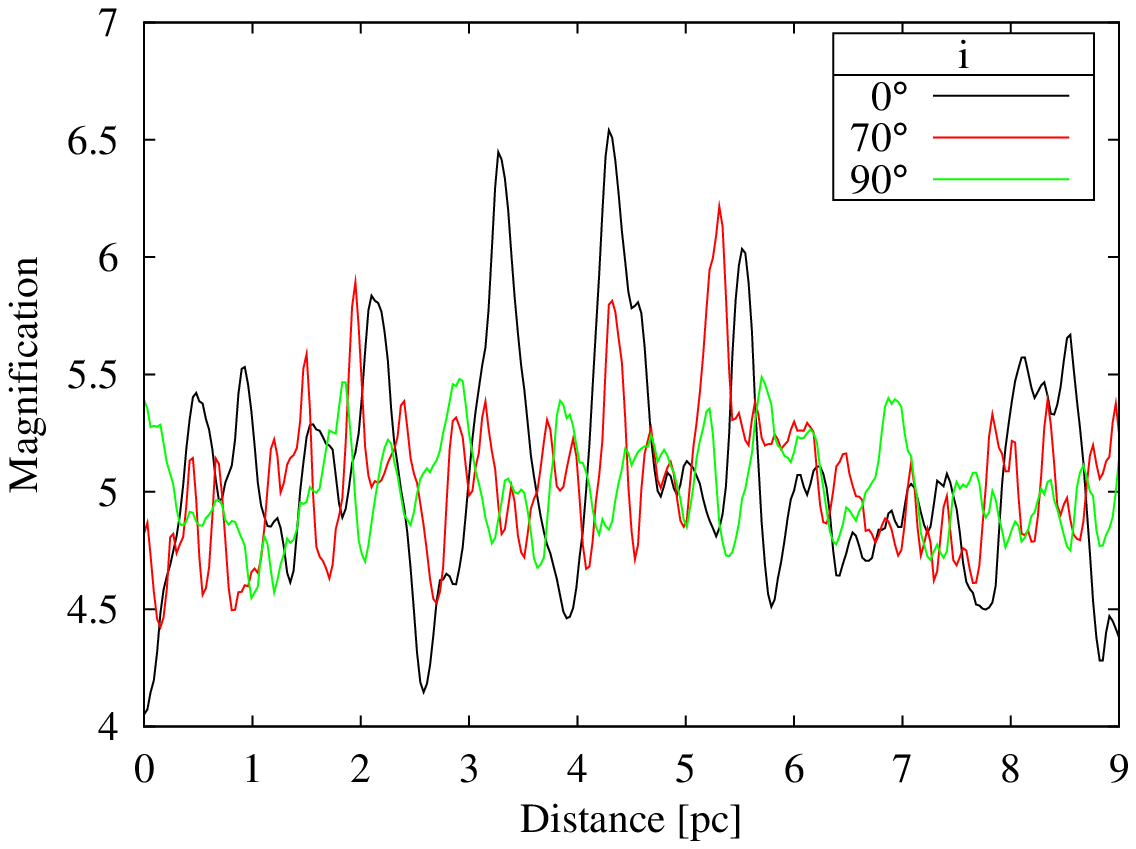}
\includegraphics[height=0.34\textwidth]{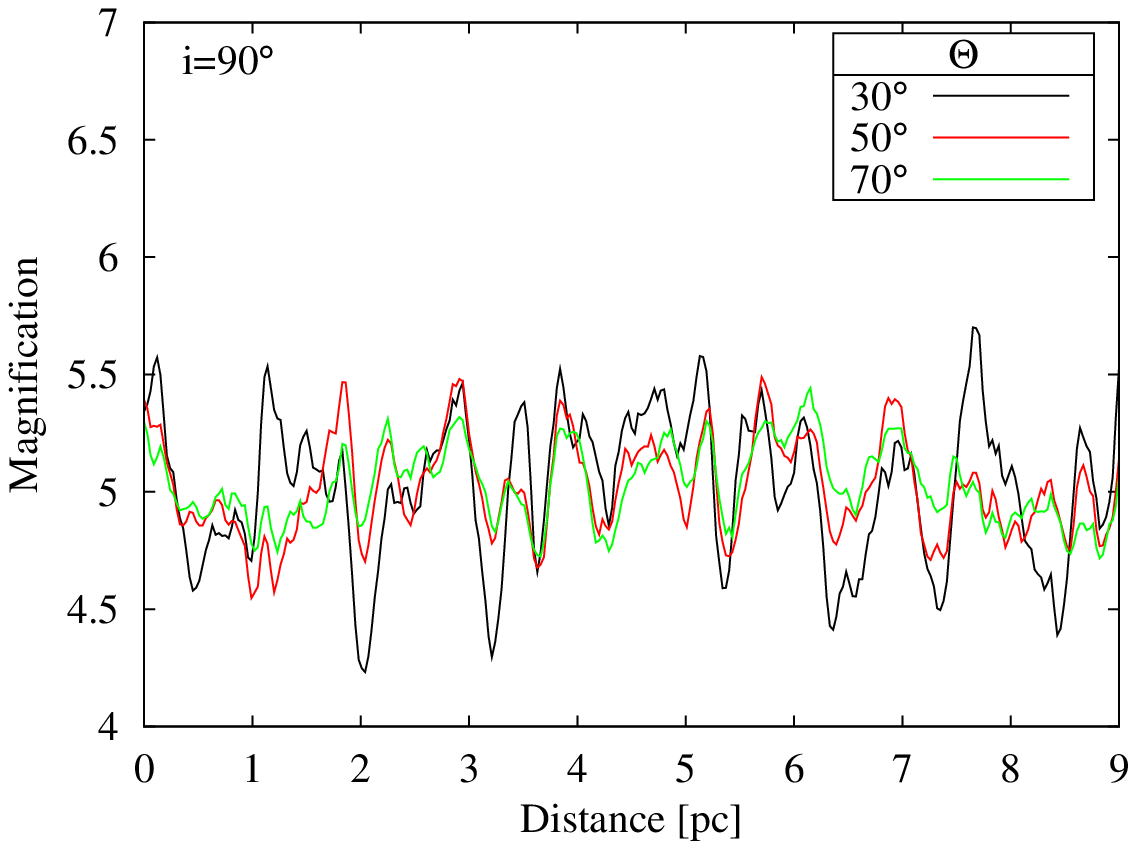}\\
\includegraphics[height=0.34\textwidth]{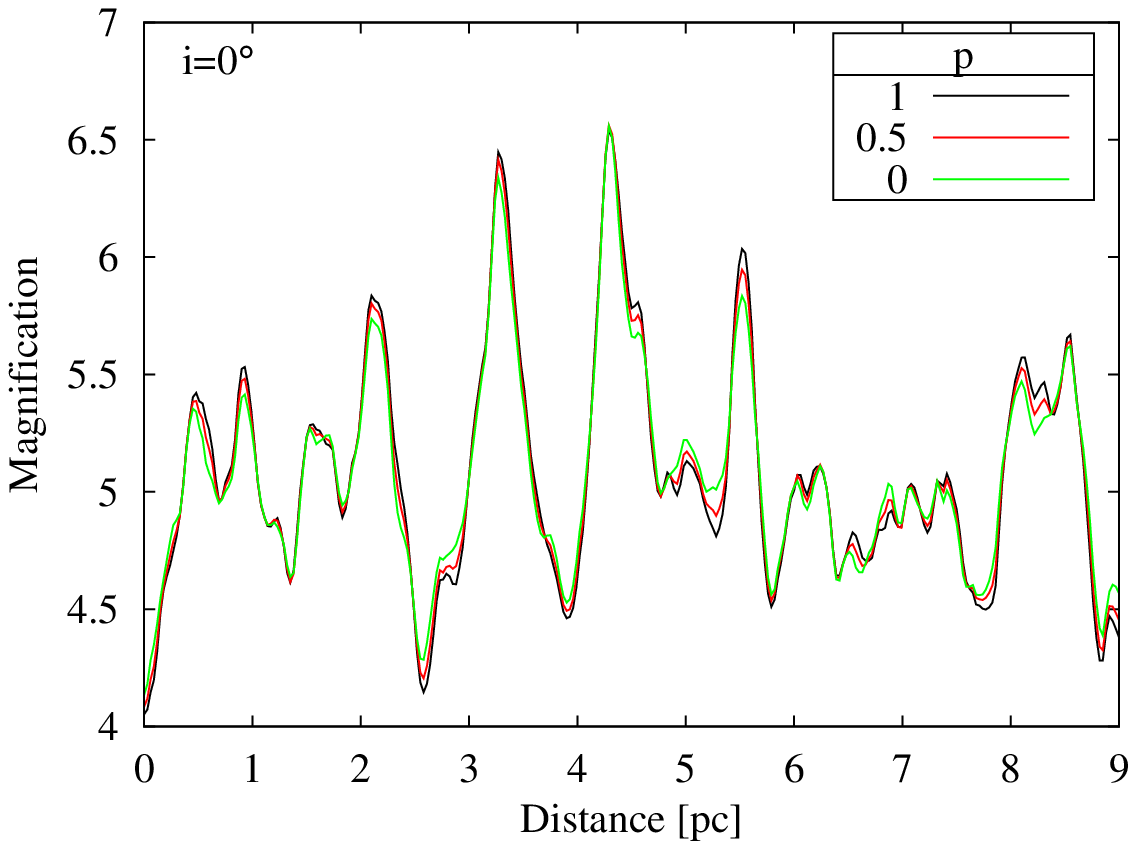}
\includegraphics[height=0.34\textwidth]{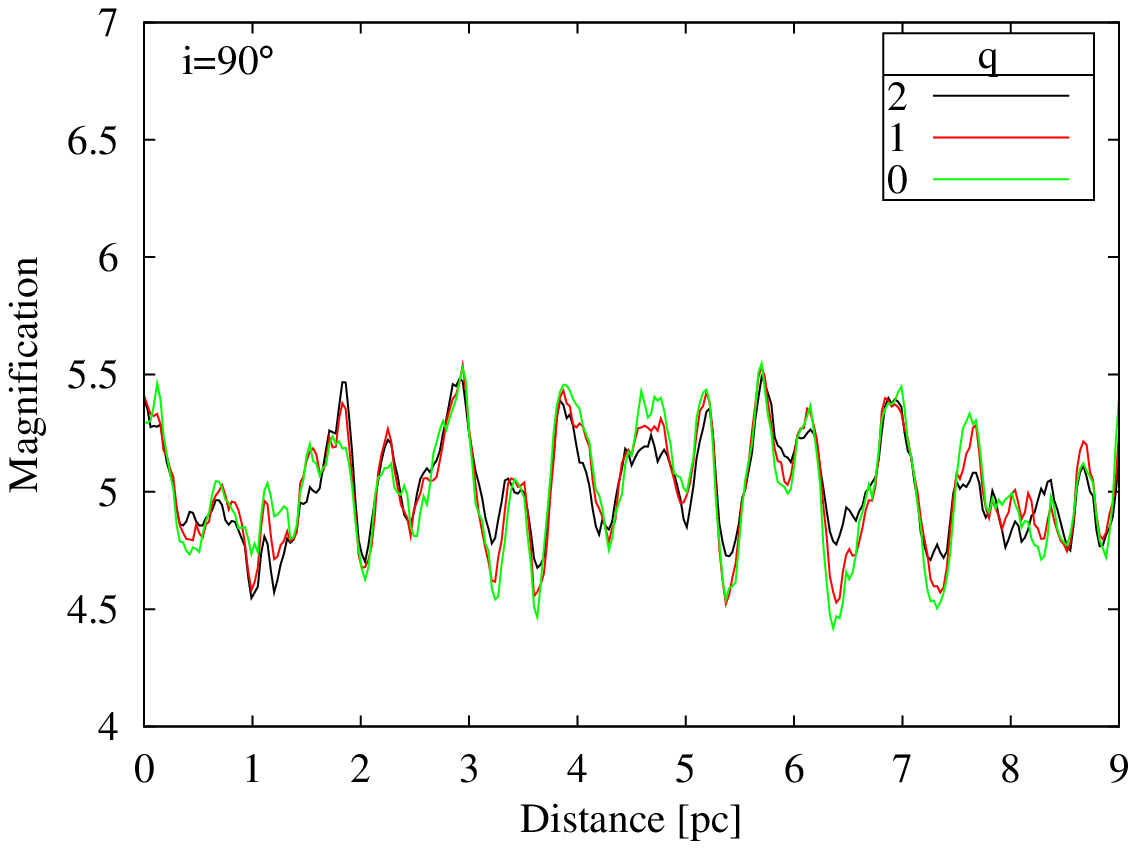}
\caption{Microlensing light curves at $1\ \mu$m for the different
parameters of the torus model. Upper left panel shows dependance on
inclination, upper right panel on half opening angle, bottom panels
on dust density distribution parameters. In all panels the black line
corresponds to the most compact dust configuration, green line for
the largest, and the red line for the intermediate.}
\label{fig:lcpar}
\end{figure*}

The parameters that determine the viewing angle, the torus size, the
dust distribution and the primary source luminosity, directly or
indirectly, all contribute to the pattern of the torus IR emission and
its apparent size at the given wavelength \citep{stalevski12}.
Therefore, each of these
parameters could affect the shape and amplitudes of microlensing light
curves. We adopted a set of parameter values as the standard ($p=1,
q=2, \Theta=50^\circ, R_{\text{out}}=3$ pc, $L=10^{12}~L_{\odot}$) and
then we varied each of these parameters, while keeping the others
constant. The resulting microlensing light curves for simulated
magnification events at $1\ \mu$m are shown in Fig.  \ref{fig:lcpar}.

The outer radius of the torus has the largest impact on the
magnification amplitude (see Fig.~\ref{fig:mamp}). Obviously, for
smaller torus sizes, the magnification will be higher. The half
opening angle ($\Theta$) is another
parameter which defines the size of torus, especially when seen
edge-on. As expected, the tori with smaller opening angle will show
larger variations under influence of microlensing.

It is evident that the viewing angle (inclination $i$) also has a
significant influence. The dust-free lines of sight (corresponding to
the type 1 AGN; $i=0^\circ$) provide a direct view of the innermost
region of very hot dust, which is obscured in the case of
dust-intercepting lines of sight (type 2 AGN; $i=70,90^\circ$). Thus,
in the former case, the larger part of the emission we see is
originating from a more compact region than in the latter case. As a
result, the amplitudes of the magnification events will be higher in
the case of type 1 objects.

The parameters $p$ and $q$ define the spatial distribution of the dust
density (Eq. \ref{eqn:dens}). The larger values correspond to the more
compact distribution of the dust. However, from the panels in the
bottom row of Fig.~\ref{fig:lcpar} we see that, although more compact
dust configuration tend to have higher amplitudes, these parameters do
not significantly influence microlensing light curves.  This is
because, for the adopted value of the primary source
luminosity ($L=10^{12}~L_{\odot}$, typical for quasars),
radiation is able to penetrate further into the dust, and thus
diminish the difference between compact and extended dust
distributions. In the case of $\sim10$ times smaller primary source
luminosities, the dust distribution parameters do have a noticable
impact.

\subsubsection{Timescales of microlensing events}
To characterize timescales of high magnification events, we use the
rise time -- the time interval between the beginning and the peak of a
microlensing event. We estimate the rise time by measuring the
distance traveled by the source, relative to the caustics, from the
beginning to the peak of HME, and divide it by the effective source
velocity. The effective source velocity $\text{V}$, i.\ e. the
velocity of the source relative to the caustics with time measured by
the observer is given by the expression \citep{kayser86}
\begin{equation}\label{eqn:ve}
   {\bf V} = {\frac{1}{1+z_{\text{l}}}}
{\frac{D_{\text{ls}}}{D_{\text{l}}} } {\bf
v}_o 
             - {\frac{1}{1+z_{\text{l}}}} { \frac{D_{\text{s}}}{D_{\text{l}}} } {\bf
v}_{\text{l}}
             + {\frac{1}{1+z_{\text{s}}}} {\bf v}_{\text{s}} .
\end{equation}
The source velocity ${\bf v}_{\text{s}}$ and the lens velocity ${\bf
v}_{\text{l}}$ are measured in the source and lens plane,
respectively, they are distance weighted, and, due to the redshifts,
translated into the observer's frame. The transverse velocity of the
observer ${\bf v}_o$ can be determined from the dipole term in the
microwave background radiation. With an amplitude of $387$ km/s
\citep[e.g.,][]{kogut93}, the observer's motion will be important for
some lenses and unimportant for others, depending on the direction
towards the source. For simplicity, we will assume here that the
direction of the observer's motion is parallel to the direction
towards the source, so the first term in the Eq.~(\ref{eqn:ve}) can
be neglected. Assuming that the peculiar velocities of the source and
lens, in their own planes, are of the same order, the last term
can be neglected as well. With these approximations, the effective
source velocity is reduced to the expression
\begin{equation}\label{eqn:ve2}
V \simeq  {\frac{1}{1+z_{\text{l}}}} {
\frac{D_{\text{s}}}{D_{\text{l}}} } v_{\text{l}} .
\end{equation}
In the case of lensed system Q2237+0305, several studies found the
transverse velocity of the lens to be in the range between
approximately $500$ and $2000$ km/s
\citep{wyithe99,kochanek04,gmerino05}. We adopted three values
in this range and calculated the corresponding effective source
velocity and rise times of HMEs. In the Table~\ref{tab:tr}, we
present estimated rise times, in years, for the two prominent HMEs
indicated in the Fig. \ref{fig:lc} with a dotted line, for different
values of lens redshift and transverse velocity of the lens.
Depending on these parameters, the obtained rise times are in range
from several decades to several hundreds of years.
\begin{table}
\centering
\caption{Rise times for HMEs, in years, calculated for different
values of lens redshift and transverse velocity of the lens. Source
is fixed at redshift $z_{\text{s}}=2$.}
\begin{tabular}{|c|c|c|c|}
\hline
\hline
& \multicolumn{3}{c|}{$v_{\text{l}}$}[km/s] \\
\cline{2-4}
$z_{\text{l}}$ & 500 & 1000 & 2000 \\
\hline
0.05 & 164 & 82 & 41 \\
\hline
0.5  & 364 & 182 & 91  \\
\hline
\hline
\end{tabular}
\label{tab:tr}
\end{table}
\subsubsection{Influence on entire IR SED}
\begin{figure}
\centering
\includegraphics[height=0.35\textwidth]{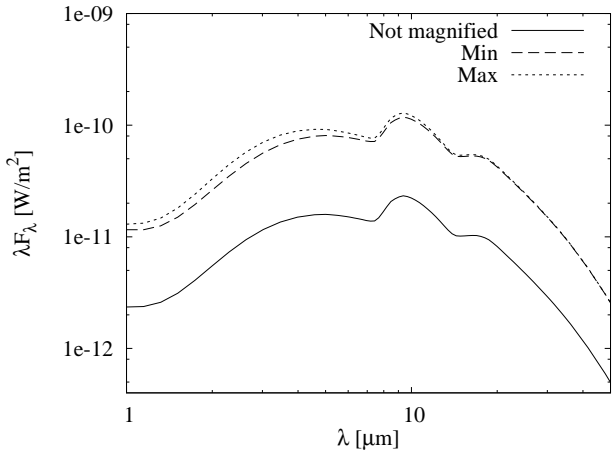}
\caption{Dusty tori SEDs in $1$--$50\ \mu$m range. Solid line:
not magnified. Dashed line: at the minumum of HME (at $\sim2.6$ pc
in the left panel Fig.~\ref{fig:lc}). Dotted line: at the maximum of
HME (at $\sim3.3$ pc in the left panel Fig.~\ref{fig:lc}). The
parameters of the torus
are the same as taken in Fig.~\ref{fig:torimg}.}
\label{fig:msed}
\end{figure}

In Fig.~\ref{fig:msed} we illustrate the influence of microlensing on
the entire dusty torus SED in the $1$--$50\ \mu$m range. From this
figure we see that there is a significant difference in the amount of
overall flux between the SEDs when microlensing is absent
(solid line) and those under the influence of microlensing (dashed
and dotted lines). However, due to the large size, the torus will
always cover a large area of a microlensing magnification pattern
and it will always be magnified by a certain factor, so
the difference between the SEDs at the beginning (dashed line) and
the peak (dotted line) of HME is marginal. Also, it is evident that
HMEs do not change the shape of SED significantly.

\section{Conclusions}
\label{sec:conc}
We investigated gravitational microlensing of AGN dusty tori in the
case of lensed quasars. The dusty torus was modeled as a clumpy
two-phase medium. The radiative transfer code \textsc{skirt} was used
to obtain SEDs and images of the tori at different wavelengths. The
ray-shooting technique has been used to calculate microlensing
magnification maps. Due to the large dimensions of dusty tori
(compared to the Einstein ring radius of the microlens in the source
plane), they must be treated as extended sources. Thus, images of the
tori were convolved with the magnification maps. We simulated
microlensing by the stars in the lens galaxy, in the case of lensed
quasars, for different configurations of the lensed system and
different values of the torus parameters, in order to estimate (a)\
amplitudes and timescales of high magnification events, and (b)\ the
influence of geometrical and physical properties of dusty tori on
light curves in the infrared domain. From our investigation, we
conclude the following.

\begin{enumerate}[(i)]
\item Despite their large size, we found that AGN
dusty torus could be significantly magnified by microlensing in some
cases. The amplitude of magnification depends on wavelenght, torus
parameters, and configuration of the lensed system.

\item The size of torus is wavelength dependent. As a consequence,
the magnification amplitude of microlensing events is also wavelength
dependent. The magnification is the highest in the near-infrared,
decreases rapidly towards the mid-infrared range, and stays almost
constant in the far-infrared part of SED.

\item As microlensing is sensitive to the size of the
source, parameters determining the geometry and the apparent size of
the torus, have a very important role. Tori with
$R_{\text{out}}\lesssim10$ pc could be appreciably
microlensed.

\hspace{10 pt} More compact dust configurations (e.g. steeper radial
density profiles) result in smaller tori and thus in higher
magnification amplitudes. However, for primary source (accretion
disk) luminosites typical for quasars ($10^{12}\ L_{\odot}$), the
influence of the dust distribution parameters is diminished, because
the radiation is able to penetrate the dust further. 

\hspace{10 pt} Tori seen at type 1 (dust-free) inclinations, which
provide a direct view of the innermost, hottest region, are more
magnified than those at type 2 (dust-intercepting) inclinations.

\item Lensed quasar systems with the lens galaxy closer to the
observer, will have higher magnification amplitudes, owing to their
larger Einstein ring radius projection on the source plane.

\item Estimated rise times, between the beginning and the peak of
HMEs, are in the range from several decades to several hundreds of
years. 
\end{enumerate}
Given such long time-scales, microlensing would hardly prove to be a
practical tool to study and constrain the properties of dusty tori,
as it is
in the case of AGN accretion disks. However, the results presented
above should be kept in mind when investigating flux ratio anomally
of lensed quasar images in different wavelength bands. In such
studies, it is important to determine the true magnification ratios
between the images, in the absence of microlensing. In principle,
this could be done by looking at the emission-line, infrared, and
radio-emitting regions of quasars, as they all should be large
enough to safely disregard microlensing effects. However, we have
shown that the infrared emission of dusty tori could be significantly
microlensed in some cases, and thus, it is a less reliable tool for
determining the ``intrinsic'' flux ratios.

\section*{Acknowledgments}

This work was supported by the Ministry of Education and Science
of the Republic of Serbia through the projects `Gravitation and
the Large Scale Structure of the Universe' (176003) and
`Astrophysical Spectroscopy of Extragalactic Objects' (176001).



\bsp

\label{lastpage}

\end{document}